\documentstyle[multicol,aps,prl,psfig,array]{revtex} 

\renewcommand{\narrowtext}{\begin{multicols}{2} \global\columnwidth20.5pc}
\def\ga{\gamma}
\def\de{\delta}
\def\ep{\epsilon}

\def\th{\theta}

\def\la{\lambda}

\def\rh{\rho}

\def\si{\sigma}

\def\ph{\phi}

\def\ps{\psi}

\def\Ph{\Phi}

\def\cD{{\cal D}}
\def\cF{{\cal F}}
\def\cL{{\cal L}}

\def\fr#1#2{{{#1} \over {#2}}}
\def\frac#1#2{\textstyle{{{#1} \over {#2}}}}
\def\pt#1{\phantom{#1}}
\def\prt{\partial}

\def\half{{\textstyle{1\over 2}}}
\def\lsim{\mathrel{\rlap{\lower4pt\hbox{\hskip1pt$\sim$}}
    \raise1pt\hbox{$<$}}}
\def\gsim{\mathrel{\rlap{\lower4pt\hbox{\hskip1pt$\sim$}}
    \raise1pt\hbox{$>$}}}
\def\ol#1{\overline{#1}}

\def\slash#1{\not\hbox{\hskip -2pt}{#1}}

\newcommand{\beq}{\begin{equation}}
\newcommand{\eeq}{\end{equation}}
\newcommand{\bea}{\begin{eqnarray}}
\newcommand{\eea}{\end{eqnarray}}
\newcommand{\rf}[1]{(\ref{#1})}

\begin{document}

\title{Supersymmetry and Lorentz Violation} 
\author{M.S.\ Berger and V.\ Alan Kosteleck\'y}
\address{Physics Department, Indiana University, 
         Bloomington, IN 47405, U.S.A.}
\date{IUHET 445, December 2001} 

\maketitle

\begin{abstract}
Supersymmetric field theories can be constructed that violate
Lorentz and CPT symmetry. 
We illustrate this with some simple examples 
related to the original Wess-Zumino model.
\end{abstract}

\bigskip

\narrowtext

A major development in fundamental theoretical physics
during the last century 
has been the understanding of the central role 
played by spacetime symmetries in nature.
Conventional spacetime symmetries,
including Lorentz and CPT invariance,
are now deeply ingrained in modern theories 
such as the standard model of particle physics and general relativity.
Recent research includes investigations 
of larger spacetime symmetries, notably supersymmetry
\cite{sw},
and of the possibility that small violations 
of conventional spacetime symmetry
could arise in an underlying theory at the Planck scale
\cite{cpt98}.

The essence of spacetime supersymmetry
is the existence of transformations between bosons and fermions 
that yield a translation operator upon anticommutation: 
\bea
&&\left[ P_\mu, P_\nu \right] 
= \left[P_\mu, Q \right] =0, \quad
\left\{Q,\ol{Q}\right\}=2\ga ^\mu P_\mu ,
\label{susy}
\eea
where 
the energy-momentum 4-vector $P_\mu$ generates spacetime translations,
the spinor $Q$ generates supersymmetry transformations,
and $\ga^\mu$ are the Dirac matrices.
Many supersymmetric Lorentz-invariant models exist.
However,
if supersymmetry is relevant to nature,
experiment suggests it must be broken.
Much of the phenomenology of supersymmetry 
conducted today is therefore
within the context of the (minimal) supersymmetric standard model
\cite{dgs}
in which supersymmetry-breaking 
but Lorentz-preserving interactions are added by hand.
This breaking is required to be soft,
in the sense that quadratic divergences continue to be avoided 
even though supersymmetry is explicitly broken. 
Soft supersymmetry-breaking terms 
are known to be superrenormalizable, 
and any nonrenormalizable terms are taken 
to be suppressed by powers of the Planck scale 
or some other large scale associated with new physics. 
Soft terms can be motivated by studies of more fundamental theories 
and could arise from spontaneous breaking of supersymmetry
\cite{hn}.
Their physical implications at low energies
can be analyzed in the framework of standard-model extensions 
including supersymmetry-breaking terms.

In a related vein, 
the physical implications of 
the breaking of conventional spacetime symmetries 
can be investigated in the context of  
a general standard-model extension 
\cite{ck}.
Its lagrangian contains terms violating Lorentz and CPT symmetry.
The nonlocal character of string theories 
offers a potential source for these terms 
\cite{kps}
and provides strong motivation 
for investigating their physical implications
at low energy.
Like the supersymmetry-breaking terms described above, 
the Lorentz- and CPT-violating terms 
could arise from spontaneous symmetry violation.
Any renormalizable terms 
would dominate in studies of low-energy physics,
since the requisite inconsistencies 
that result from treating the nonlocal underlying theory
as a local field theory emerge at 
a high-energy scale determined by the Planck mass
\cite{kle}.

In this work,
we consider an issue unaddressed in the literature:
the existence of fully supersymmetric theories
incorporating violation of Lorentz symmetry.
For simplicity,
we restrict attention here to global (rigid) supersymmetry
and consider only renormalizable models conserving energy and momentum.
\it A priori, \rm
even the existence of such theories is unclear,
and in fact  
we find the simultaneous presence of supersymmetry
and Lorentz violation provides a strong restriction
on possible models.

For definiteness,
we perform the analysis in the context of 
the Wess-Zumino model in four spacetime dimensions
\cite{wz}.
This model involves a scalar field $A$, a pseudoscalar $B$,
a Majorana fermion $\ps$, an auxiliary scalar
$F$, and an auxiliary pseudoscalar $G$.
The associated lagrangian ${\cL}_{\rm WZ}$
can be written as
\bea
{\cL}_{\rm WZ}&&=
{\cL}_{\rm k.e.} + {\cL}_{m} + {\cL}_{g} ,
\label{wesszumino}
\eea
where the kinetic terms are 
\bea
{\cL}_{\rm k.e.}&=&{\half}\prt _\mu A\prt ^\mu A 
+{\half}\prt _\mu B\prt ^\mu B 
\nonumber\\
&&
+{\half}i\ol{\ps}\slash{\prt}\ps+{\half}F^2+{\half}G^2,
\label{ke}
\eea
and the mass and interaction terms are 
\bea
{\cL}_{m} &=&
m\left (-{\half}\ol{\ps}\ps+AF+BG\right ),
\nonumber \\
{\cL}_{g} &=&
\fr g {\sqrt 2}
\left [F(A^2-B^2)+2GAB-\ol{\ps}(A-i\ga _5B)\ps\right ].
\label{mg}
\eea
To facilitate contact with existing studies of 
the Lorentz-violating standard-model extension,
we adopt the conventions of Ref.\ \cite{ck}
throughout this work.

Consider the special Lorentz-violating 
but CPT-preserving extension 
of the Wess-Zumino model given by the lagrangian
\bea
&&{\cL} = {\cL}_{\rm WZ} + {\cL}_{\rm Lorentz},
\label{lagrangian}
\eea
where 
\bea
{\cL}_{\rm Lorentz}&=&
k_{\mu \nu}\prt ^\mu A\prt ^\nu A 
+k_{\mu \nu}\prt ^\mu B\prt ^\nu B 
\nonumber \\
&&
+{\half}k_{\mu \nu}k^\mu _{\pt{\mu}\rh}(\prt ^\nu A\prt ^\rh A
+\prt ^\nu B\prt ^\rh B)
\nonumber \\
&&
+{\half}ik_{\mu \nu}\ol{\ps}\ga^\mu \prt ^\nu \ps.
\label{lorentz}
\eea
In this equation,
$k_{\mu\nu}$ is a real, symmetric, traceless, 
and dimensionless coefficient 
determining the magnitude of Lorentz violation,
which for simplicity
we suppose to be small in the chosen observer inertial frame.
The coefficient $k_{\mu\nu}$
transforms as a 2-tensor under observer Lorentz transformations
but is unaffected by particle Lorentz transformations
\cite{ck}.

Direct calculation reveals that the model 
given by Eqs.\ \rf{lagrangian} and \rf{lorentz} 
is invariant up to a total derivative
under the following set of modified 
infinitesimal supersymmetry transformations:
\bea
\de A&=&\ol{\ep}\ps,
\nonumber \\
\de B&=&i\ol{\ep}\ga _5 \ps, 
\nonumber \\
\de \ps&=&-(i\slash{\prt}+ik_{\mu \nu}\ga ^\mu\prt ^\nu)
(A+i\ga _5B)\ep+(F+i\ga _5G)\ep, 
\nonumber \\
\de F &=& -\bar{\ep}(i\slash{\prt}+ik_{\mu \nu}
\ga^\mu \prt ^\nu)\ps, 
\nonumber \\
\de G &=& \bar{\ep}(\ga _5\slash{\prt}
+k_{\mu \nu}\ga_5\ga^\mu \prt ^\nu)\ps.
\label{transformation}
\eea
In this equation,
$\ep$ is a constant Majorana spinor. 
These transformation laws are observer covariant,
so the supersymmetry is independent of the observer inertial frame.
However,
the presence of $k_{\mu\nu}$
implies the transformations are realized differently on particles
with different orientations and boosts,
as is to be expected in a theory with Lorentz violation.
Note that the usual Wess-Zumino transformations
are recovered in the limit $k_{\mu\nu} \to 0$.

The commutator of two supersymmetry transformations \rf{transformation}
yields
\bea
\left[ \de_1,\de_2 \right] 
= 2 i \ol\ep_1 \ga^\mu \ep_2 \prt_\mu 
+ 2 i k_{\mu\nu}\ol \ep_1 \ga^\mu \ep_2 \prt^\nu,
\eea
which involves the generator of translations.
A modified supersymmetry algebra therefore exists.
A superspace realization of this superalgebra
is discussed below.

The lagrangian \rf{lagrangian}
thus provides an explicit example of an interacting model 
with both exact supersymmetry and Lorentz violation.
We know of no other supersymmetric, CPT-preserving,
and Lorentz-violating extension of the minimal Wess-Zumino multiplet.
The possible supersymmetry transformation laws are strongly restricted
by various factors,
including the linearity in $\ep$ and the fields,
the small number of physical Lorentz-violating terms for Majorana spinors,
the properties of coefficients for Lorentz violation,
and the requirement of closure of the induced supersymmetry algebra.

The presence of $k_{\mu\nu}$ in 
the supersymmetric transformation forces a relationship 
on the coefficients for Lorentz violation
in Eq.\ \rf{lorentz}.
This is analogous to the common mass and common couplings 
that are a standard consequence of supersymmetric theories. 
Without the supersymmetry,
each of the five terms in Eq.\ \rf{lorentz}
could have different coefficients,
a variety that is reflected 
in the form of the general Lorentz-violating standard-model extension.
Physical consequences of
the relationship among the coefficients in Eq.\ \rf{lorentz}
are to be expected.
For example,
the fermionic propagator is
\bea
&&
i S_F(p) = \fr i {p_\mu (\ga^\mu + k_{\mu \nu}\ga ^\nu) - m}.
\eea
Rationalizing the denominator of this propagator gives
\bea
&&
i S_F(p) = i \fr {p_\mu (\ga^\mu + k_{\mu \nu}\ga ^\nu) + m}
{ p^2+2 p^\mu p^\nu k_{\mu \nu}
+ k_{\mu \rh}k^\rh_{\pt{\rh}\nu}p^\mu p^\nu},
\eea
using the symmetry of $k_{\mu \nu}$.
Consequently,
the scalar and fermionic propagators have the same structure.
We therefore anticipate divergence cancellations 
and nonrenormalization theorems generalizing the usual results.

Note also that the Lorentz violation of the theory is physical.
The interactions eliminate the possibility of 
a trivializing field redefinition
\cite{ck}.
For example,
if $\cL_g$ were absent,
then for suitable $\tilde k_{\mu\nu}$ a simultaneous field redefinition 
$f(x) \to \exp(\tilde k_{\mu\nu}x^\mu\prt^\nu)f(x)$
of all fields in the supermultiplet
would eliminate the terms in $\cL_{\rm Lorentz}$ 
while leaving unaffected the mass terms $\cL_m$.
However,
with $\cL_g$ present the same field redefinition
merely replaces $\cL_{\rm Lorentz}$ 
with $x$-dependent Lorentz-violating interactions.

The Lorentz-violating CPT-preserving model 
\rf{lagrangian} 
can be described in a superfield formulation. 
Define 
\bea
&&\ph = \frac 1 {\sqrt 2} (A + i B),
\quad
\cF = \frac 1 {\sqrt 2} (F - i G).
\eea
In terms of these complex scalars,
the left-chiral superfield appropriate for the model  
\rf{lagrangian} is
\bea
\Ph(x,\th) &=& 
\ph (x) + \sqrt 2 \ol \th \ps_L (x) 
+ \half \ol \th (1 - \ga_5) \th \cF (x) 
\nonumber\\
&&
+\half i \ol \th \ga_5 \ga^\mu \th 
(\prt_\mu + k_{\mu\nu} \prt^\nu)\ph (x) 
\nonumber\\
&&
- \frac i {\sqrt 2} \ol \th \th \ol\th 
(\slash\prt + k_{\mu\nu}\ga^\mu\prt^\nu) \ps_L (x)
\nonumber\\
&&
- \frac 1 8 (\ol\th\th)^2 (\prt_\mu + k_{\mu\nu}\prt^\nu)^2 \ph (x).
\label{superfield}
\eea
Here, 
the subscript $L$ denotes projection 
with $\half (1 - \ga_5)$.
The lagrangian \rf{lagrangian}
can then be expressed as 
\bea
\cL = \Ph^* \Ph\vert _D
+ \left( \half m \Ph^2\vert _F
+ \frac 13 g \Ph^3\vert_F + {\rm h.c.}\right),
\label{sfieldlag}
\eea
where the symbols $\vert_D$ and $\vert_F$
refer to projections onto the $D$- and $F$-type components 
of the (holomorphic) functions of $\Ph (x, \th)$.
The theory can therefore be represented as an action in superspace.

A superspace realization $Q$
of the supersymmetry generators  
can be obtained via a coset-space construction 
\cite{ss}.
For a supersymmetry transformation on $\Ph (x, \th)$
generated as
$\de_Q \Ph (x,\th) = - i \ol\ep Q \Ph(x,\th)$,
the form of $Q$ is found to be 
\bea
Q&=& i \prt_{\ol \th} - \ga^\mu \th \prt_\mu
- k_{\mu\nu} \ga^\mu \th \prt^\nu.
\label{q}
\eea
This induces the supersymmetry transformations \rf{transformation} 
on the component fields in $\Ph(x,\th)$.

The superalgebra generated by $Q$
and $P_\mu = i \prt_\mu$ is 
\bea
&&\left [P_\mu, Q \right ] =0,
\quad
\left \{Q,\ol{Q}\right \}
=2\ga ^\mu P_\mu  + 2 k_{\mu\nu}\ga^\mu P^\nu.
\label{superalg}
\eea
By virtue of the Lorentz violation,
manifest through the presence of $k_{\mu\nu}$,
this superalgebra lies outside the usual list 
of possible supersymmetric extensions of the Poincar\'e,
de Sitter, or conformal algebras
\cite{hls}.
It appears feasible and would be of interest to
obtain a general classification of such superalgebras
allowing for the possibility of Lorentz violation.

As a more technical remark,
we observe that a superfield covariant derivative $\cD$ can be introduced 
in analogy with the usual case:
\bea
\cD&=& i \prt_{\ol \th} 
+ \ga^\mu \th \prt_\mu
+ k_{\mu\nu} \ga^\mu \th \prt^\nu.
\label{d}
\eea
It obeys
\bea
\left \{\cD,\ol{\cD}\right \}&=&-2\ga ^\mu P_\mu 
- 2 k_{\mu\nu} \ga^\mu P^\nu,
\label{dalg}
\eea
and has vanishing anticommutators with $Q$, $\ol Q$.
The form of Eq.\ \rf{dalg} implies the geometry of superspace is changed
in that the torsion is modified by the presence of Lorentz violation.
The right projection $\half(1 + \ga_5)\cD$ 
defines a left-chiral coordinate $x_+^\mu$
through the condition ${\cD}_R x_+ = 0$:
\bea
&&x_+^\mu = 
x^\mu 
+ \half i \ol \th \ga_5\ga^\mu\th 
+ \half i k^{\mu\nu} \ol \th \ga_5\ga_\nu\th.
\label{chiralcoord}
\eea
In terms of $x_+^\mu$,
the left-chiral superfield \rf{superfield} 
takes the simpler form
\bea
\Ph(x,\th) &=& 
\ph(x_+) + \sqrt 2 \ol \th_R \ps_L(x_+)
+ \ol \th_R \th_L \cF (x_+),
\label{xplussuperfield}
\eea
and is annihilated by $\cD_R$:
\beq
{\cD}_R \Ph(x_+, \th) = 0.
\eeq

The form \rf{superalg} of the superalgebra
involves the generator $P^\mu$ of translations.
A conserved canonical energy-momentum tensor $\th^{\mu\nu}$
can be constructed,
and $P^\mu$ is then recovered as
the spatial integral of the components $\th^{0\mu}$
\cite{ck,kle}.
The presence of derivative couplings in $\cL_{\rm Lorentz}$
means that care is required in physical interpretation
because the physically propagating supermultiplet 
differs by a field redefinition 
from the superfield components of $\Ph(x,\th)$.
Also,
the 4-momenta for one-particle states
obey modified dispersion laws.
However,
\bea
\left [Q,P^2 \right ] =0,
\eea
so the eigenvalues of $P^2$ must be
the same for members of the supermultiplet.
Since the superpotential containing the mass and coupling terms
is unaffected by the Lorentz violation, 
analogues should exist for
various conventional results on supersymmetry breaking 
\cite{susyb}.
Note also that a supersymmetry current can be obtained
because the supersymmetry is 
a continuous global symmetry of the lagrangian.
The existence of the superfield formulation 
implies a corresponding supercurrent superfield
can be constructed.

In the context of spontaneous Lorentz violation 
in an underlying covariant string field theory,
the coefficients $k_{\mu \nu}$ would be related to 
one or more vacuum expectation values of Lorentz vector or tensor fields
\cite{kps}. 
The form of the transformations \rf{transformation} 
then suggests that the supersymmetry 
must be realized in a nonlinear fashion
in the underlying string field theory,
since the coefficients $k_{\mu\nu}$ would be associated
with dynamical fields.
Note also that,
even if a \it linear \rm
supersymmetry in the underlying string theory
breaks along with Lorentz symmetry,
the model \rf{lagrangian} demonstrates that
an exact linear supersymmetry 
could still exist in the effective low-energy theory.
 
We next consider the more difficult challenge 
of constructing a CPT-violating extension 
of the Wess-Zumino model.
It is a famous result of quantum field theory 
that a local Lorentz-invariant theory 
preserves the combination CPT 
\cite{blp}. 
However,
if Lorentz invariance is abandoned,
one can consider the addition of a CPT-odd component 
to $\cL_{\rm WZ}$,
\bea
&&{\cL} = {\cL}_{\rm WZ} + {\cL}_{\rm CPT},
\label{lagrangian2}
\eea
where
\bea
{\cL}_{\rm CPT} &=&
k_\mu(A\prt ^\mu B-B\prt ^\mu A)
+{\half}k^2(A^2+B^2)
\nonumber \\
&&
-{\half}k_\mu \ol{\ps}\ga _5\ga^\mu \ps .
\label{cpt}
\eea
Here, the CPT violation is controlled by $k_\mu$,
which is a real coefficient of mass dimension one
transforming as a vector under observer Lorentz transformations
but as a scalar under particle Lorentz transformations.
The terms \rf{cpt} respect C but violate P or T, 
giving an overall CPT violation.
The appearance of the terms with coefficient $k^2 = k_\mu k^\mu$ 
represents a mass renormalization  
varying with the particle boost and orientation.
This is necessary for the existence of the supersymmetry below,
except in the special case of lightlike $k_\mu$.

The model \rf{lagrangian2} transforms into a total derivative
under the infinitesimal supersymmetry transformations
\bea
\de A&=&\ol{\ep}\ps, \nonumber \\
\de B&=&i\ol{\ep}\ga _5 \ps, \nonumber \\
\de \ps&=&-(i\slash{\prt}+\ga _5\slash{k})(A+i\ga _5B)\ep
+(F+i\ga _5G)\ep, \nonumber \\
\de F &=& -\bar{\ep}(i\slash{\prt}-\ga _5\slash{k})\ps,
\nonumber \\
\de G &=& \bar{\ep}(\ga _5\slash{\prt}+i\slash{k})\ps.
\label{transformation2}
\eea
The uniqueness of this supersymmetry 
can be established on dimensional grounds.
Note that it acts differently 
on the left-chiral multiplet and its conjugate,
for example,
\bea
\de \ps_L=(-i\slash{\prt}+\slash{k})(A+iB)\ep_R
+(F-iG)\ep_L, \nonumber \\
\de \ps_R=(-i\slash{\prt}-\slash{k})(A-iB)\ep_L
+(F+iG)\ep_R.
\eea

The terms \rf{cpt} would emerge from $\cL_{\rm WZ}$ 
under the field redefinition
\bea
\ps &\to&  e^{-i\ga_5 k\cdot x}\ps,
\nonumber\\
(\ph,\cF)&\to& e^{i k \cdot x}(\ph,\cF).
\label{fredef}
\eea
The components of the left-chiral multiplet 
and its conjugate 
are therefore shifted by opposite position-dependent phases.
The mass and coupling terms in $\cL_g$
would acquire CPT- and Lorentz-violating position-dependent coefficients
under the field redefinition,
so if energy-momentum is to be conserved 
they would need to be added afterwards.
However,
they are then inconsistent with the supersymmetry \rf{transformation2}.
The same is true of P-odd mass or coupling terms,
such as the combination
$(i \ol{\ps}\ga _5\ps+2AG-2BF)$.
In the absence of noninvariant couplings,
the field redefinition implies that  
the CPT and Lorentz violation in Eq.\ \rf{cpt}
is unphysical. 

Acting on the components of the left-chiral multiplet,
the commutator of two supersymmetry transformations 
\rf{transformation2} gives
\bea
\left[ \de_1,\de_2 \right]\vert_{\rm left}
= 2 i \ol \ep_1 \ga^\mu \ep_2 \prt_\mu - 2 k_\mu \ol \ep_1 \ga^\mu \ep_2,
\label{lcm}
\eea
which again involves the generator of translations.
The last term is a special consequence of Lorentz violation,
absent in the conventional spacetime superalgebras
but allowed here because $k_\mu$ has mass dimension one.
However,
the commutator of two supersymmetry transformations 
on the right-chiral multiplet yields instead 
\bea
\left[ \de_1,\de_2 \right]\vert_{\rm right}
= 2 i \ol \ep_1 \ga^\mu \ep_2 \prt_\mu + 2 k_\mu \ol \ep_1 \ga^\mu \ep_2.
\label{rcm}
\eea
The relative sign change in the last term complicates 
a superspace construction.
It has features reminiscent of central charges 
for conventional extended supersymmetry.
It would be interesting to obtain 
an explicit superspace formulation of this model 
with a differential realization 
of the supersymmetry transformations
that reproduces the intertwined relations 
\rf{lcm} and \rf{rcm}.
In any case,
however,
there would be an obstacle to 
construction of an invariant superpotential
involving the usual $F$-type terms:
the $F$ term no longer transforms as a total derivative
under a supersymmetry transformation,
as follows from Eq.\ \rf{transformation2}.

Although it lies beyond our present scope,
it would be of interest to investigate the possibility
of Lorentz-violating models with extended supersymmetry.
Certainly,
$N=1$ models similar to those in Eqs.\ \rf{lagrangian} and \rf{lagrangian2}
but involving several supermultiplets
appear straightforward to construct.
The presence of several multiplets might permit physical CPT violation,
although more general field redefinitions that mix fields between
multiplets would need to be considered.
It may also be useful to allow for variant multiplets
in constructing Lorentz-violating models.
For example,
the scalar $\ph$ can be regarded as the dual of an antisymmetric 2-tensor,
for which the extra spacetime indices might permit
distinct Lorentz-violating couplings.
Note also that various renormalizable Lorentz- and CPT-violating terms 
exist that are unused in the model discussed above,
including
$(A^2 \prt B \pm B^2 \prt A)$,
$\ph\ol\ps \ga_5\ga^\mu\ps$,
and $\ol{\ps}\si ^{\mu \nu}\prt ^\la \ps$.
It is unclear whether there is 
a supersymmetric role for these terms,
which are associated with dimensionless 
coefficients for Lorentz and CPT violation
carrying one or three spacetime indices.

One can extend the considerations discussed 
here to other representations of supersymmetry. 
For example,
the vector supermultiplet has a Lorentz-violating generalization,
so a supersymmetric Lorentz-violating extension of quantum electrodynamics 
should exist.
Similarly,
it appears feasible to construct
a supersymmetric Lorentz-violating standard-model extension,
in which case potentially realistic models could be obtained 
by including soft supersymmetry-breaking terms.
These soft terms would include 
Lorentz-violating dimension-three operators
of the types discussed in Ref.\ \cite{ck}.
In the context of supergravity models,
the scale $m$ of the soft terms is often related to 
the scale $M_s$ of supersymmetry breaking 
by $m\sim M_s^{(1+n)}/M_{P}^n$
for some integer $n>0$. 
Generalizing the results here to local supersymmetry
and local Lorentz violation 
might therefore eventually uncover determining relationships 
among the scale of Lorentz violation,
the scale of supersymmetry breaking,
and the underlying Planck scale. 

This work was supported in part 
by DOE grant DE-FG02-91ER40661.

\end{multicols}
\end{document}